\documentclass[10pt,twocolumn,letterpaper]{article}

\usepackage{cvpr}
\usepackage{times}
\usepackage{epsfig}
\usepackage{graphicx}
\usepackage{amsmath}
\usepackage{amssymb}
\usepackage{multirow}
\usepackage[subrefformat=parens]{subcaption}

\usepackage[breaklinks=true,bookmarks=false]{hyperref}

\cvprfinalcopy % *** Uncomment this line for the final submission

\def\cvprPaperID{****} % *** Enter the CVPR Paper ID here

\begin{document}

%%%%%%%%% TITLE
\title{Stroke Correspondence by Labeling Closed Areas}

\author{Ryoma Miyauchi\\
JAIST\\
Ishikawa, JAPAN
\and
Tsukasa Fukusato\\
The University of Tokyo\\
Tokyo, JAPAN
\and
Haoran Xie\thanks{xie@jaist.ac.jp}\\
JAIST\\
Ishikawa, JAPAN
\and
Kazunori Miyata\\
JAIST\\
Ishikawa, JAPAN
}

\maketitle
%\thispagestyle{empty}

%%%%%%%%% ABSTRACT
\begin{abstract} 
Constructing stroke correspondences between keyframes is one of the most important processes in the production pipeline of hand-drawn inbetweening frames. This process requires time-consuming manual work imposing a tremendous burden on the animators. We propose a method to estimate stroke correspondences between raster character images (keyframes) without vectorization processes. 
First, the proposed system separates the closed areas in each keyframe and estimates the correspondences between closed areas by using the characteristics of shape, depth, and closed area connection. 
Second, the proposed system estimates stroke correspondences from the estimated closed area correspondences. 
We demonstrate the effectiveness of our method by performing a user study and comparing the proposed system with conventional approaches.

\end{abstract}

%%%%%%%%% BODY TEXT
\section{Introduction}
Inbetweening generation from keyframes plays an important role in 2D animation production. The inbetweening frames are interpolated between hand-drawn images. For example, approximately 800 keyframes and  5000 to 7000 inbetweenings are required for a 30-minutes animation. In conventional animation, keyframes have to be manually vectorized to unified strokes from the original drawings, where strokes are generally rough and uneven. This process is commonly tedious and time-consuming even for skilled animators. 

To solve this issue, the inbetweening generation approaches were examined using two consecutive keyframes~\cite{baxter2006latent, boris2015vector, di2001automatic, fukusato2016active, fukusato2021view, fujita2017canvas, Furusawa2014QR, Marek2018, sederberg1992physically, sederberg19932, yang2017context}. However, these approaches were mainly based on vectorized drawings and manual stroke correspondence between the keyframes. It remains a challeng to automatically obtain the stroke correspondence from original drawings without vectorization.

\begin{figure}[t]
\centering
\includegraphics[clip,width=90mm]{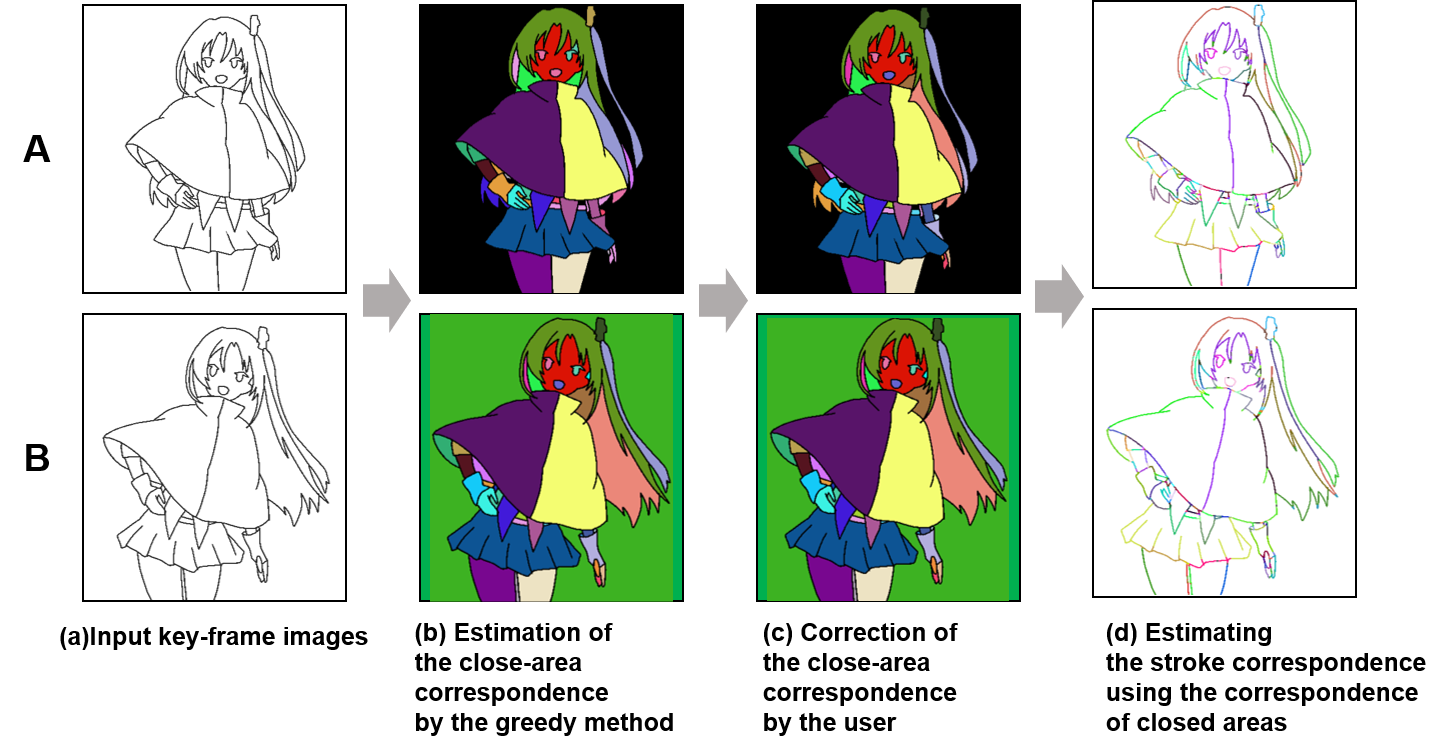}
\caption{ The work flow of the proposed stroke correspondence support system:
Paired closed areas are given the same color.
}
\label{greedy1}
\end{figure}

In raster keyframes, each stroke plays an important role in representing object contours and depth information. Given this, we consider the drawn strokes of keyframes as boundary lines of the separated closed areas. Therefore, we propose a method to interactively estimate stroke correspondences between two raster keyframes by using closed area correspondences, as shown in Figure~\ref{greedy1}.

The proposed system can help users avoid the tedious work of redrawing the vector lines. 
This system can also automatically generate inbetweening frames. 
To verify the feasibility of the proposed system, we calculate the accuracy of the estimated closed area correspondence using several examples of animation cuts. In addition, we measure the amount of user corrections until the closed area correspondence is completely determined. 
We confirm the visual qualities of stroke correspondences by importing the estimated stroke correspondence to commercial inbetweening products, such as Cacani~\cite{cacani2021}.

\begin{figure*}[t]
\centering
\includegraphics[clip,width=150mm]{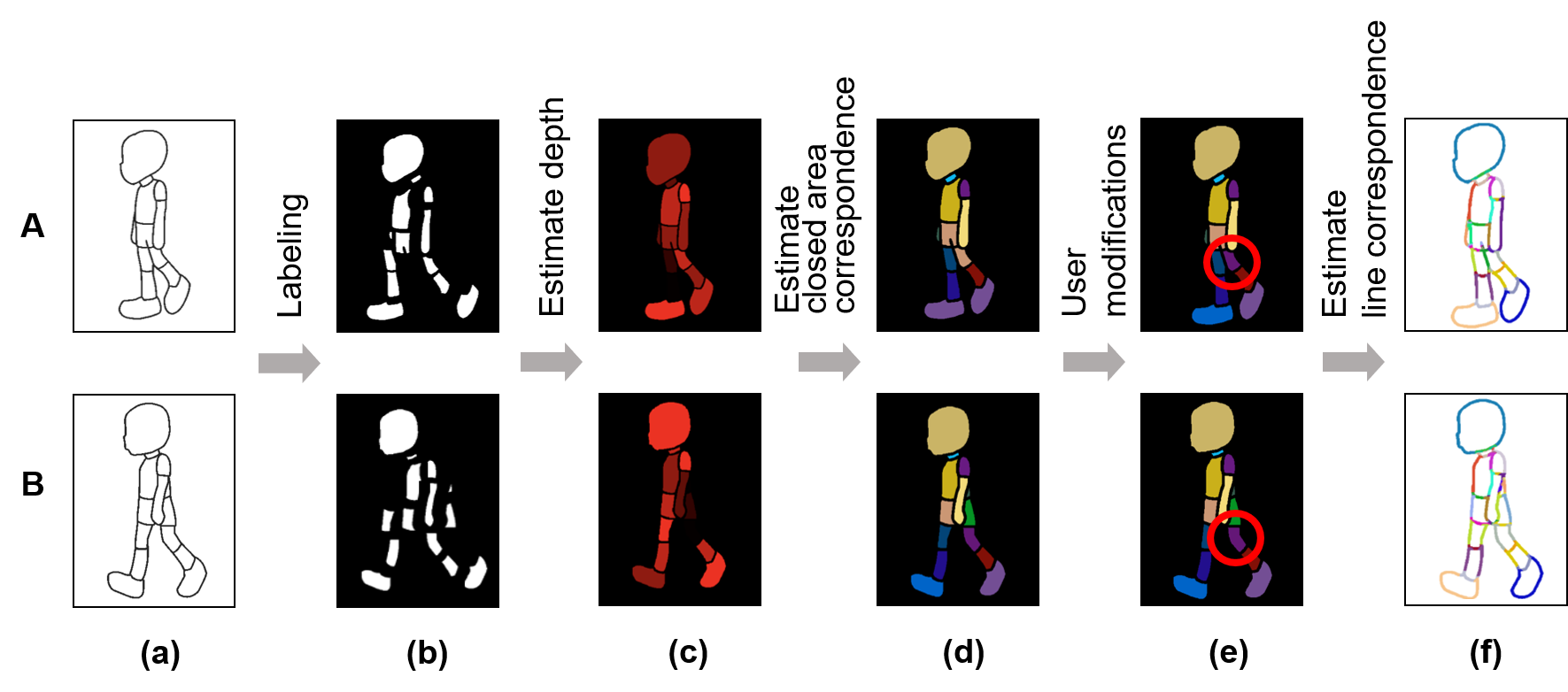}
\caption{ System overview. The process includes (a)~inputting two line drawings (keyframes) $A$ and $B$,
(b)~separating each keyframe into regions (closed areas),
(c)~estimating depth relationships between the closed areas,
(d)~estimating closed area correspondences between keyframes by using a greedy algorithm, 
(e)~modifying corresponding closed area manually,
and (f)~estimating stroke correspondences using the closed area correspondences.
}
\label{outline}
\end{figure*}

\section{Related Work}
\subsection{Stroke Correspondence}
Many techniques have been proposed to automatically generate inbetweening frames from keyframes, which are composed of strokes (vector lines)~\cite{baxter2006latent, boris2015vector, di2001automatic, fukusato2016active, Furusawa2014QR, Marek2018, sederberg1992physically, sederberg19932}. In general, stroke correspondences between keyframes are determined by imposing restrictions, such as drawing orders~\cite{burtnyk1975computer, reeves1981inbetweening}. However, in many cases, a keyframe is composed of hundreds of strokes, and users invest a lot of time and effort to determine the correspondence. 
In addition, several studies on automatic methods have focused on a narrower range of target strokes, such as boundary lines~\cite{Juan2006reusing, baxter2009compatible}, but they are difficult to determine inner strokes in closed area. Hence, recent studies discuss how to construct stroke correspondence.
%[5][6]

In studies of Fujita et al.\cite{fujita2017canvas, fujita2019}, two keyframes are projected onto a three-dimensional virtual canvas. Several strokes set by the users in each keyframe will overlay each other when transforming the shape of the canvas, and the correspondence is estimated by combinatorial optimization based on distance.
Whited et al.~\cite{white2010betweenit} construct a connection graph based on a  relationship between the keyframes, estimating the correspondence by performing a depth-first search on the graph. However, this process fails when
storkes are occluded in more complex scenes, or change in shapes and numbers of strokes between keyframes take place.

%AL8
Yang et al.~\cite{yang2017context} utilize characteristics of neighboring strokes (e.g., stroke shapes), and propose a method to create a correspondence between keyframes by using a greedy algorithm. This system enables users to manually specify several constraints and handle the estimated results (correspondence). 
In addition, Yang et al.~\cite{yang2018ftp} extend Yang's method~\cite{yang2017context} to establish characteristics of stroke connections, and improve the estimation accuracy. 
However, with these methods, the users must manually redraw keyframes with vector lines by tracing over the original character image in advance. 

Therefore, we focus on closed areas of character images (i.e., the character parts) with a method to directly estimate stroke correspondence from character images. Note that the manual vector redrawing process can be skipped, hence users can concentrate on designing movements of each stroke.

\subsection{Closed area Correspondence}
Constructing closed area correspondences between two raster images has been thoroughly investigated~\cite{yamaguchi2015region}. In the field of reference-based colorization of line drawings, Sato et al.~\cite{sato2014reference} utilize the positional and connection relationships of closed areas in the keyframes to construct their correspondence. Maejima et al.~\cite{maejima2019graph} propose a method to (i)~estimate the corresponding closed areas in the reference keyframes and (ii)~ color target inbetweenings based on (i) estmation.
While these approaches are suitable for colorization tasks, they are unsuitable for inbetweening generation because they cannot make vertex correspondences to each pair of corresponding strokes. 
Consequently, we combine closed area and stroke-based methods that presume to be the optimum method. With this proposed frameworks, we can directly design inbetweenings from two keyframes without vectorization steps.

\section{Proposed method}
In this section, we introduce a framework to construct stroke correspondences. First, we separated two input images (keyframes) into closed areas while removing noises. Then, we estimate closed area correspondences between keyframes by using a greedy algorithm. Note that the user can manually correct the correspondence using the correction interface. Finally, the system automatically estimates stroke correspondences by referring to the estimated closed area correspondence.

\subsection{Overview}
Figure~\ref{outline} shows the overview of the proposed system to estimate the stroke correspondence between the two continuous keyframes. Closed areas in the two input keyframes are labeled, and the depth context relationship of the closed areas is estimated, followed by a greedy algorithm to estimate all closed area correspondences between the two keyframes. Finally, users correct the closed area correspondence if they consider it is necessary, and the area correspondence is updated from the corrected results. Stroke correspondence relationship between the two input keyframes is determined from the estimated closed area correspondence.

\subsection{Labeling Process}
\begin{figure}[t]
\centering
\includegraphics[clip,width=0.7\linewidth]{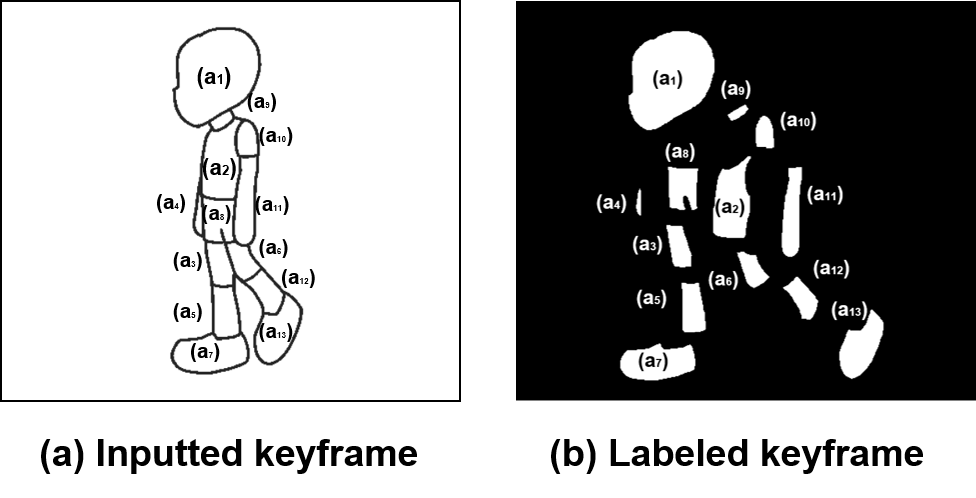}
\caption{An example of labeling. (a)~an input keyframe and (b)~separated regions (closed areas) using OpenCV library.}
\label{label}
\end{figure}

We first scan images $A$ and $B$ (keyframes) and import them into the PC. The color images are converted to grayscale to perform the labeling process with OpenCV, and then a median filter in kernel size $5$ removes fine noises and repairs broken strokes.
Subsequently, closed areas in the keyframes are labeled with threshold binarizing $220$ pixels, which achieved the most stable results in  multiple experiments. 
The labeled closed area of keyframe A is $\{A|(a_1,a_2,\cdots,a_n)\}$, and the closed area of  keyframe B is $\{B|(b_1,b_2,\cdots,b_m)\}$, as shown in Figure~\ref{label}.

%++++++++++++++++++++++++++
%++++++++++++++++++++++++++

Information such as assigning color, area, coordinates of the centroid, and the label and angle of the centroid in the adjacent closed area must be obtained from closed area $a_i(i=1,2,\cdots,n)$ and $ b_i(i=1,2,\cdots,m)$ .
At this time, the closed area that touches the greatest proportion of the screen's outer circumferences is estimated as the background.

\subsection {Depth Context of Closed Area}
We extract the intersections in the input keyframe using OpenCV, and draw a circle with radius $r$ placing the intersections as a center point. The proposed system estimates the depth context of each closed area from the ratio of the closed areas in those circles. Radius $r$ is regarded as the square root of ($2*$(screen size-background label size) $/10000$) so that the area of the circle is $1/100$ the area of the character on the screen. 

%++++++++++++++++++++++++++
\begin{figure}[t]
\centering
\includegraphics[clip,width=\linewidth]{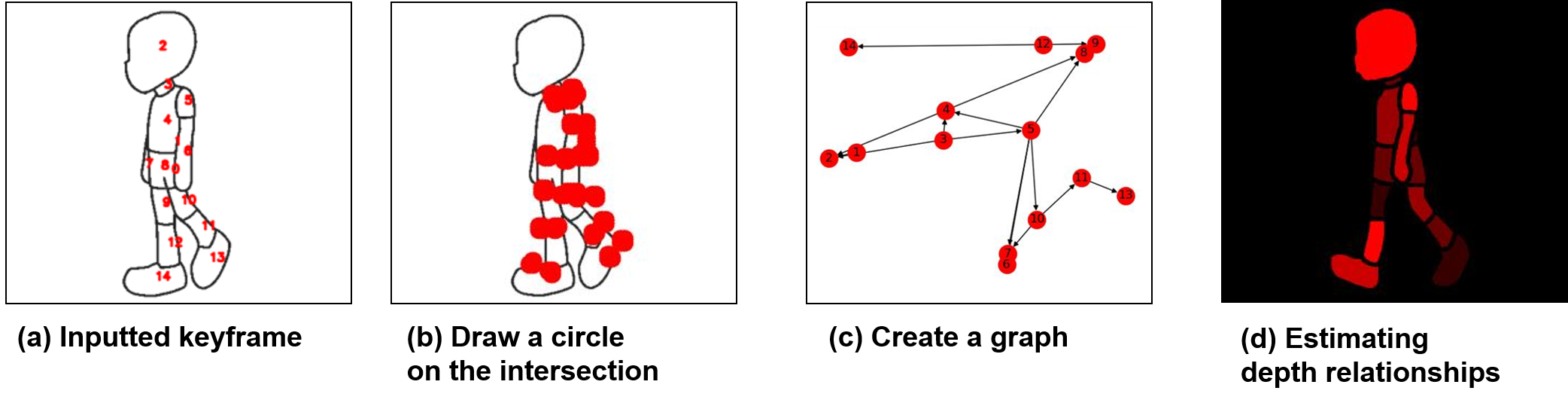}
\caption{ Depth estimation. (a)~assigning label IDs $i \in \{1,\cdots N\}$ of the closed areas, 
(b)~extracting intersections and connection relationships between closed areas, (c)~constructing a weighted graph, and (d)~estimating depth relationships.
}
\label{depth2}
\end{figure}
%++++++++++++++++++++++++++

As a result, the information of the adjacent closed area in the same keyframe and its own depth context relationship are obtained as for the information of the closed areas $a_i, (i = 1,2,\cdots,n)$ and $b_i(i=1,2,\cdots,m)$.
%(Figure~5)

\subsection{Greedy Algorithm}

Our proposed system focuses on estimating all closed area correspondence as the target problem. This problem can be solved with the estimation of the closed area correspondence near the seed pair, which is considered to be the pair among all combinations as sub-problem. Therefore, we use a greedy algorithm to solve this problem and seek local optimum solutions.

In our algorithm, the score of the closed area near the seed pair is calculated and updated. The first seed pair is the starting point of the greedy algorithm. Then, the pair with the highest score is considered as a new seed pair, and the score is updated iteratively. We repeat the calculation until the scores of all closed areas are updated at least once. Finally, we swap the first and the second keyframes to improve accuracy and update the scores accordingly.

To determine the seed pair, it is necessary to calculate the seed score by quantifying the similarity of the closed areas. The closed area of the labeled keyframe A is $\{A|(a_1,a_2,\cdots,a_n)\}$. The closed area of the keyframel B is $\{B|(b_1,b_2,\cdots,b_m)\}$. The formula for calculating the seed score of the closed area is $a_i (i=1,2,\cdots,n)$ ,as follows.
\begin{multline}
\label{SeedScore}
SeedScore(a_i,b_j)\\
=max\{N(a_i,b_j)S(a_i,b_j)\}(j=1,2,\cdots,m)
\end{multline}

\noindent
where $N(a_i,b_j)$ denotes the calculated value by matching feature points between closed areas $a_i$ and $b_j$ using AKAZE~\cite{akaze}. $S(a_i,b_j)$ denotes the area ratio of closed area $a_i$ to $b_j$. Note that $S(a_i,b_j)$ is $a_i/b_j$ when $a_i> b_j$, and $b_j/a_i$ when $a_i<b_j$.

The seed score is calculated for all combinations of $a_i$ and $b_j$, and the combination with the highest seed score is considered to be the first seed pair $(a_{seed},b_{seed})$. For determining the seed, the seed score of the other keyframe’s closed area is obtained as a tentative score for the information of closed areas $a_i(i=1,2,\cdots,n)$ and $b_j(j=1,2,\cdots,m)$.

To estimate correspondence near seed pair, the seed pair $(a_{seed},b_{seed})$ updated their scores using the positional relationship and depth context relative to the seed pair.

\subsection {Closed Area Association}
Let $\{A_{seed}|(a_1,a_2,\cdots,a)\}$ be the set of closed area near $a_{seed}$, and let $\{B_{seed}|(b_1,b_2,\cdots,j)\}$ be the set of closed area near $b_{seed}$.  
We calculate the score $R(a_{seed},b_{seed},a_i,b_j)$ as the relationship between $A_{seed}$ and $a_{seed}$. The formula for calculating this relationship is given below.
\begin{multline}
\label{R}
R(a_{seed},b_{seed},a_i,b_j)
\\=(1+((a-AngleDef(a_{seed},b_{seed},a_i,b_j))/a)^2)
\\(1+(Situation(a_{seed},b_{seed},a_i,b_j)/s))
\end{multline}

\noindent
where $a$ and $s$ are constant values. $AngleDef = abs(\theta_a-\theta_b)$ denotes that the absolute value of the difference between angle $a_{seed}$ and $a_i$ is $\theta_a$; the angle between $b_{seed}$ and $b_j$ is $\theta_b$. $Situation$ denotes a step function that compares the context of $a_{seed}$ and $a_i$ with the context of $b_{seed}$ and $b_j$, outputting 1 if they are similar and 0 if they are different. The formula for calculating the score of the closed area $a_i (i=1,2,\cdots,n)$ is as follows (The highest score $b_j$ becomes the complement of $a_i$).
\begin{multline}
\label{Score}
Score(a_i,b_j)
\\=PS(N(a_i,b_j)S(a_i,b_j)R(a_{seed},b_{seed},a_i,b_j))
\\(j=1,2,\cdots,m)
\end{multline}

\noindent
where $N(a_i,b_j)$ and $S(a_i,b_j)$ are similar to the seed score in Equation (\ref{SeedScore}), and $R(a_{seed},b_{seed},a_i,b_j)$ in Equation (3.3) is the score (Equation \ref{R}) which is the correspondence between closed areas $a_i$ and $b_j$). The tentative score obtained when calculating the seed score is $PS$. 

At this step, the paired closed areas  $ a_i $  and  $ b_j $  are determined as a new seed pair, the correspondence in the neighboring seed pair is estimated, and all the closed area scores are updated. This loop performs as many times as the number of the closed areas in the keyframes, and the highest score becomes the first seed pair. Then the score is updated again by swapping keyframes A and B. In order to prevent an infinite loop, a pair that has already been treated as a seed pair will not be retreated if an alternative seed pair can be selected. 

\begin{figure}[t]
\centering
\includegraphics[clip,width=\linewidth]{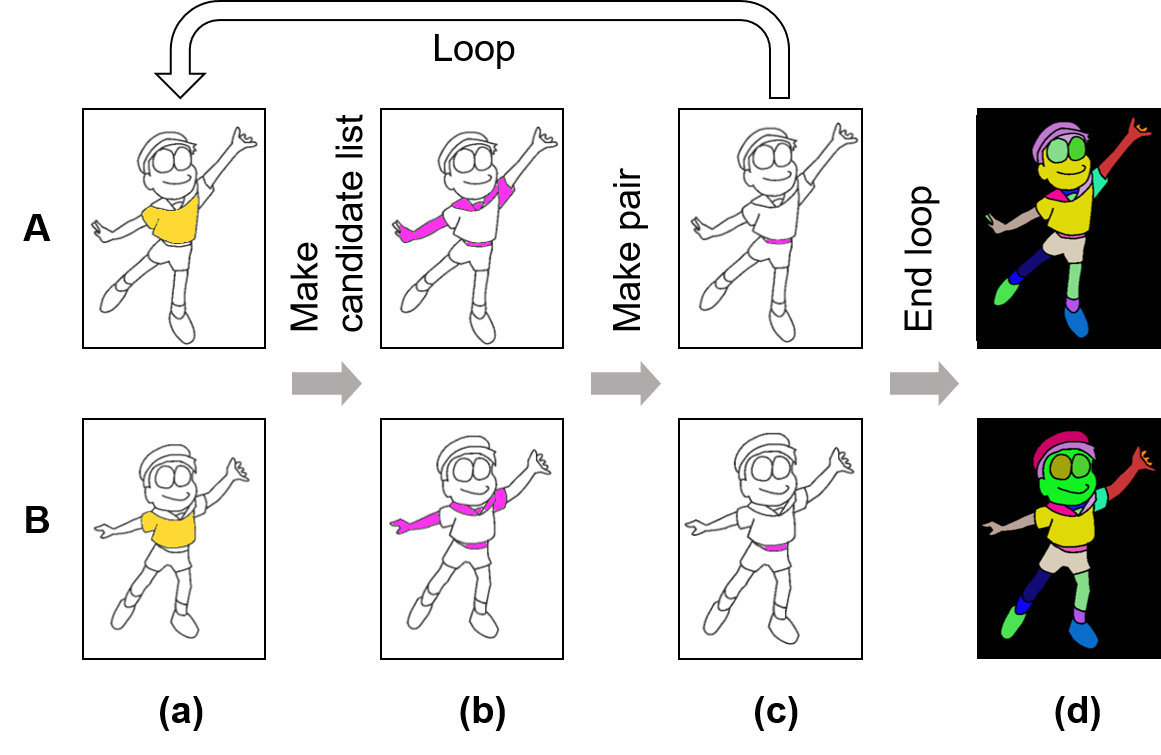}
\caption{The proposed process for estimating closed area correspondences involves (a)~identifying a seed pair (yellow), (b)~updating the scores of the closed areas neighboring the seed pair (pink), (c)~selecting a pair with the highest score (pink) and replacing the initial seed pair with it, and (d)~repeating until the correspondence is deemed complete.}
\label{greedy}
\end{figure}

Figure~\ref{greedy} shows the work process of the greedy algorithm, that estimates closed area correspondence. 

\subsection {User Correction}
When errors are found in the closed area correspondence estimated by the greedy algorithm, users can manually correct them with our interface. Based on the correction, the system re-calculates the scores and the correspondence. Theses corrections can also improve the accuracy of stroke correspondences.

\subsection {Stroke Correspondence}
\begin{figure}[t]
\centering
\includegraphics[clip,width=0.85\linewidth]{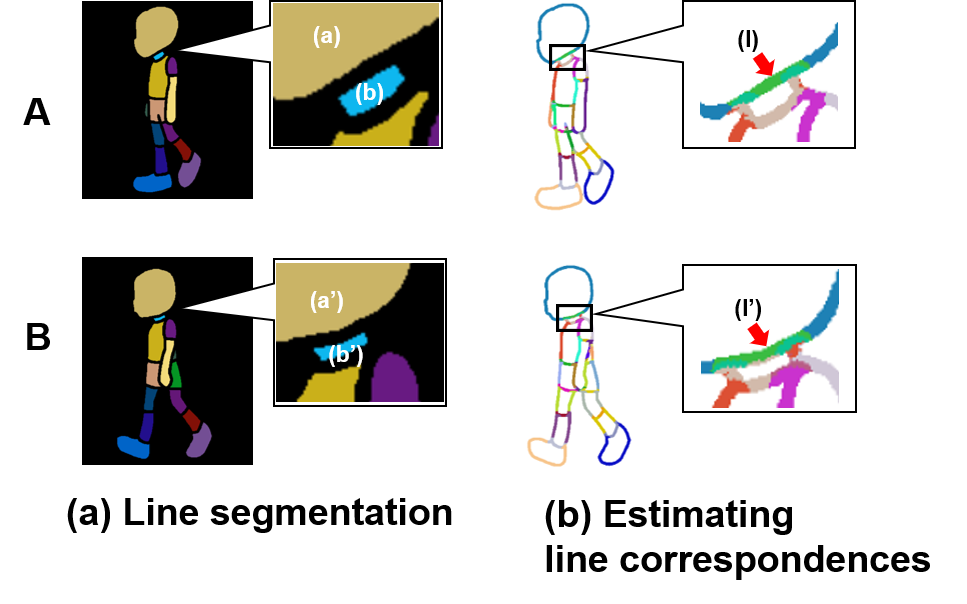}
\caption{Stroke correspondence estimation involves (a)~separating each closed areas into strokes, and (b)~estimating stroke correspondence using closed area correspondences.
}
\label{stroke}
\end{figure}

Boundary lines of the closed areas in keyframes are regarded as strokes. Therefore, we construct stroke correspondences by adopting the estimated closed area correspondences.
Figure~\ref{stroke} shows the process of the stroke correspondence estimation. For example, when (1)~two closed areas in keyframe A ($a$ and $b$) are separated by one stroke $l$ (green) and (2)~two closed areas in keyframe B ($a'$ and $b'$) are separated by one stroke $l'$ (green), we can easily determine that stroke $l$ corresponds to stroke $l'$. Therefore, we find correspondences between strokes that separated two or more closed areas.

This method for determining stroke correspondence between two consecutive keyframes suggests the possibility of avoiding the trouble of redrawing the keyframes in vector lines, which is a requirement in previous work.  In addition, matching accuracy can be improved by considering not only the corresponding relationships of the position and connection, but also the depth relationships when determining closed area correspondence.

%++++++++++++++++++++++++++++

\vspace{2mm}
\section{User Study}
We implemented the proposed approach on a desktop computer with Intel i7-10700 CPU 2.90GHz and GeForce RTX 2070 SUPER GPU. The purpose of the user study is to verify the feasibility of estimating stroke correspondence by using closed area labeling and to determine whether the burden on the user is reduced, compared to the existing method. To verify user's burden reduction  compared to the existing method, we used Cacani~\cite{cacani2021} to construct stroke correspondences. We investigated the effectiveness of the proposed system through questionnaires. To determine the feasibility of the system, we measured the time required to carry out the work.

\subsection{Comparison Study}
Our main objective is not to automatically construct the correspondence but rather provide a good starting point for manual editing. Hence, we incorporate a manual editing step to modify closed area correspondences in the keyframes. This process is similar to Cacani's procedure: (1)~initializing the correspondence using an inbetweening generator~\cite{burtnyk1975computer} and (2)~modifying it manually. Therefore, this experiment adopted as eveluation criteria both the accuracy of closed area correspondences estimation (without user correction), and the number of user corrections until the closed area completely matched.

\vspace{1mm}
\subsubsection{Experimental Method}
To verify the estimation qualities, we estimated closed area correspondences with the following three methods:

\vspace{2mm}
\begin{itemize}
    \item $SCD$:\hspace{2mm}shape, connection and depth relationship.
    \item $SC$:\hspace{5mm}shape and connection relationship.
    \item $S$:\hspace{8mm}shape only.
\end{itemize}
\vspace{2mm}

We measured the matching accuracy of closed area correspondences and stroke correspondences between the two consecutive keyframes (without user correction). The accuracy of the closed area correspondence between the input keyframes A and B is calculated by the following equation.
\begin{multline}
\label{41}
AreaMatch(A,B)
\\=\{(A_n + B_n) - MissmatchArea \}/(A_n + B_n)
\end{multline}

\noindent
where $A_n$ is the number of closed areas in keyframe A, $ B_n $ is the number of closed areas in keyframe B, and $MissmatchArea$ is the difference of closed areas with changed pairs, between the correspondence before user correction and the complete correspondence after user correction. We also measured the number of manual corrections until the closed areas completely matched. 

Similarly, the accuracy of stroke correspondences is calculated by the following equation.
\begin{multline}
\label{42}
LineMatch(A,B)
\\=\{(A_n + B_n) - MissmatchLine \}/(A_n + B_n)
\end{multline}

\noindent
where $ A_n $ is the number of strokes of keyframe A split by the author,  $ B_n $ is the number of strokes of the keyframe B split by the author, and $MissmatchLine$ is the number of different pair strokes due to the different results from Cacani versus the proposed method. Note that no pair result from the proposed method is counted as $MissmatchLine$.

In addition, we imported the estimated stroke correspondence of the two consecutive keyframes (with user correction step) into Cacani to identify (1)~the visual qualities of inbetweening frames and (2)~the feasibility of stroke correspondence estimation using our method.

\vspace{1mm}
\subsubsection{Case Study}
We used four types of animation cuts from online video resources: 

\vspace{2mm}
\begin{itemize}
\setlength{\leftskip}{2mm}
    \item[$C_a$:] cut with little movement.
    \item[$C_b$:] cut with large movement.
    \item[$C_c$:] cut with little movement and change in the front-back relationship.
    \item[$C_d$:] cut with large movement and change in the depth context relationship.
\end{itemize}
\vspace{2mm}

\subsection {System Evaluation}
We conducted system evaluation with 9 graduate students aged 20--40 years working in information systems in Japan (6 males and 3 females) including experienced animators. We verified the burden felt by the users when determining the corresponding closed areas and estimating the corresponding strokes using the proposed method, versus determining the corresponding strokes using Cacani.
%in 20s and 30s

\subsubsection {Experimental method}
\begin{figure}[t]
\centerline{\includegraphics[clip,width=80mm]{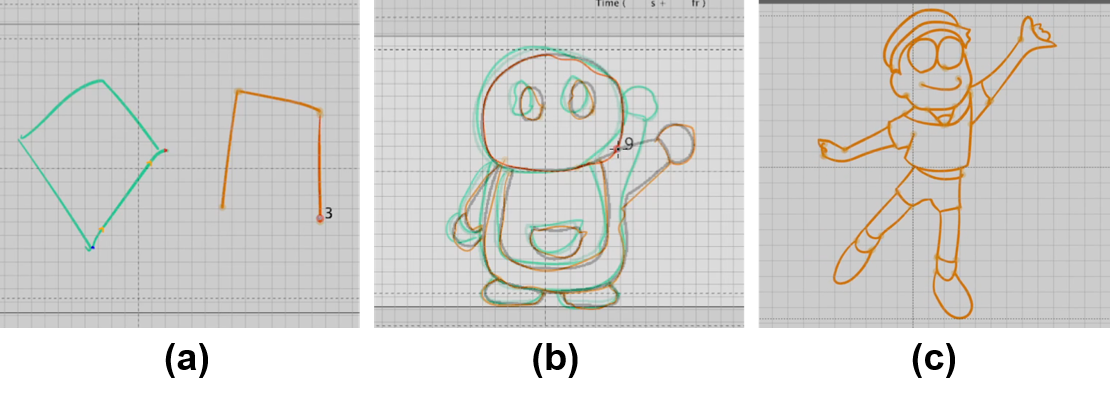}}
\caption{(a, b)~Simple drawings for training task and (c)~more complex keyframes created by tracing an actual image from broadcasted animation.}
\label{ex21}
\end{figure}

The participants were asked to determine the corresponding closed areas and the corresponding strokes from the two keyframes using the proposed method and Cacani. Figure~\ref{ex21} shows keyframes used in the experiment. 

%++++++++++++++++++++++++++++
%++++++++++++++++++++++++++++

First, we took 10 minutes to explain how to operate Cacani, we then asked the participants to practice using simple figures (see Figure~\ref{ex21}(7a and 7b)). We limited the practice to 20 minutes to determine whether the participants fully understood Cacani. 
Next, we asked the participants to construct stroke correspondences of complex characters (see Figure~\ref{ex21}(7c)). Note that all participants used the same work environment.

At the end of the construction process, we examined the feasibility of the proposed method through questionnaires. In addition to the survey on their burden, we also investigated whether the results were satisfied and whether the operation was easy to understand. 
%
%+++++++++++++++++++++++++
\begin{figure*}[t]
\begin{minipage}[b]{0.52\linewidth}
\includegraphics[clip,width=90mm]{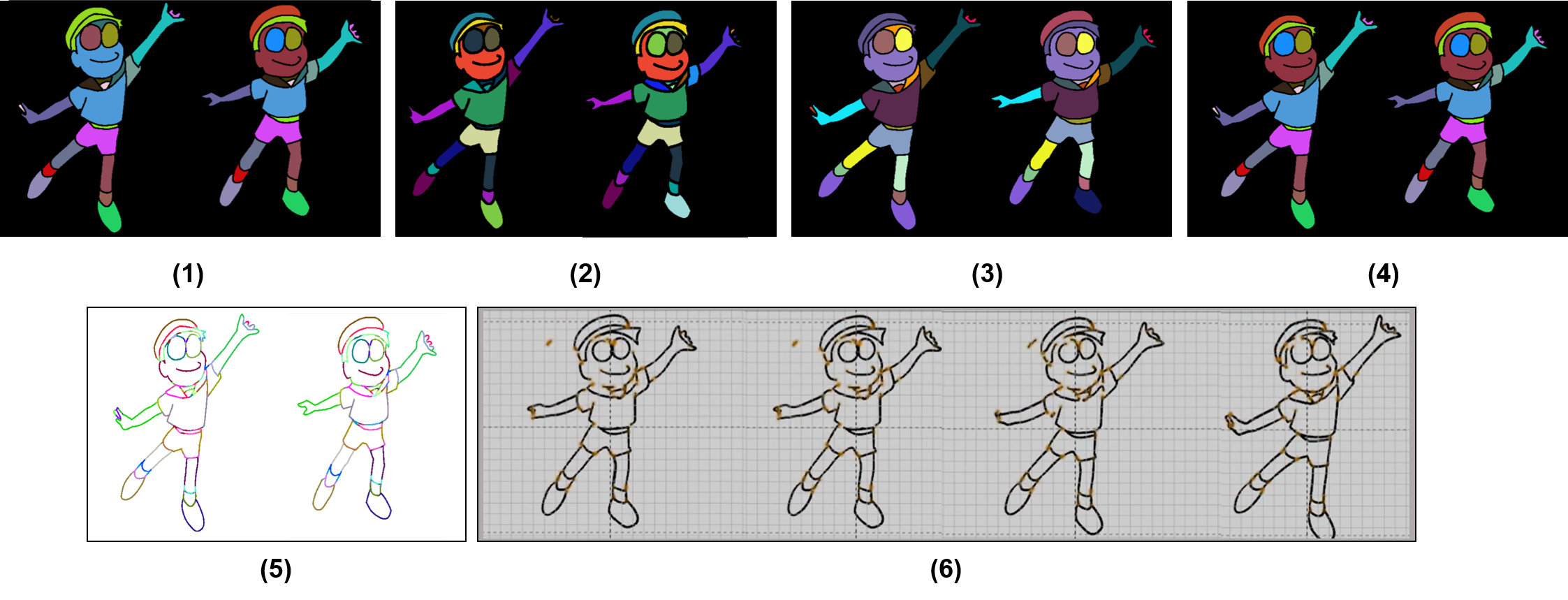}
\subcaption{$C_a$}
\label{e11}
\end{minipage}
\begin{minipage}[b]{0.52\linewidth}
\includegraphics[clip,width=90mm]{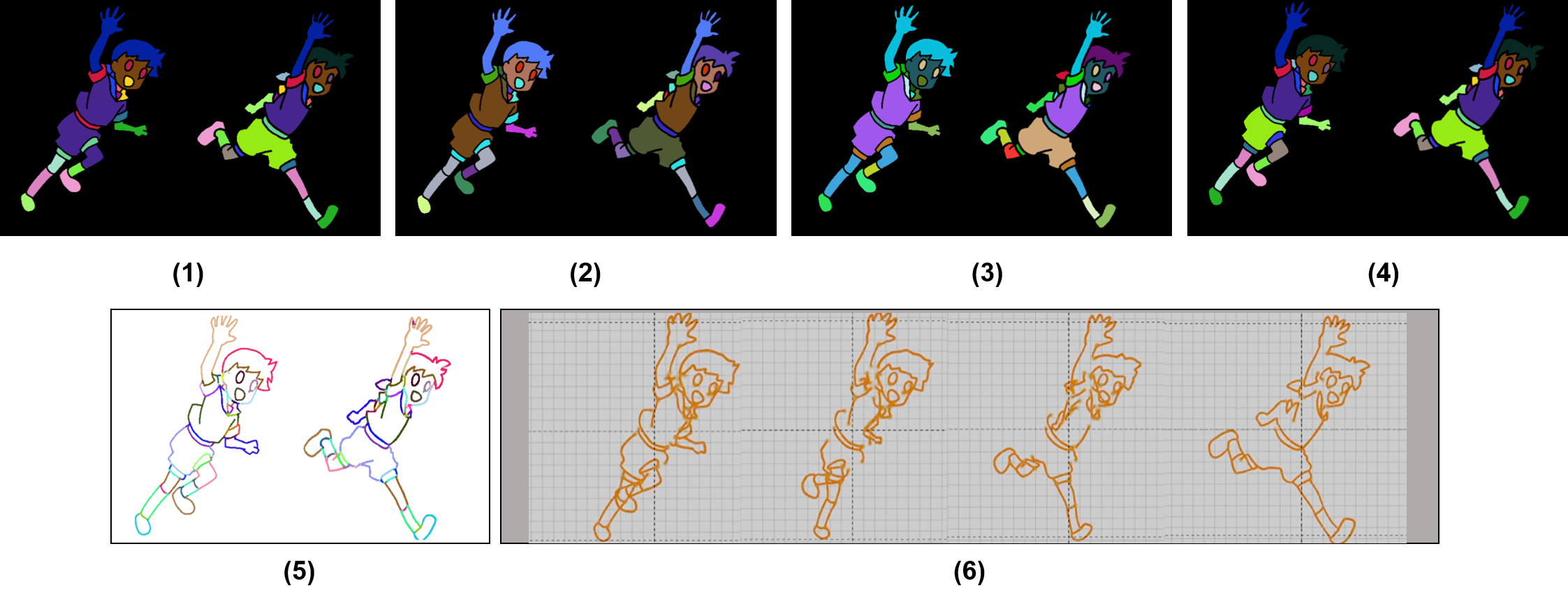}
\subcaption{$C_b$}
\label{e12}
\end{minipage}

\begin{minipage}[b]{0.52\linewidth}
\includegraphics[clip,width=90mm]{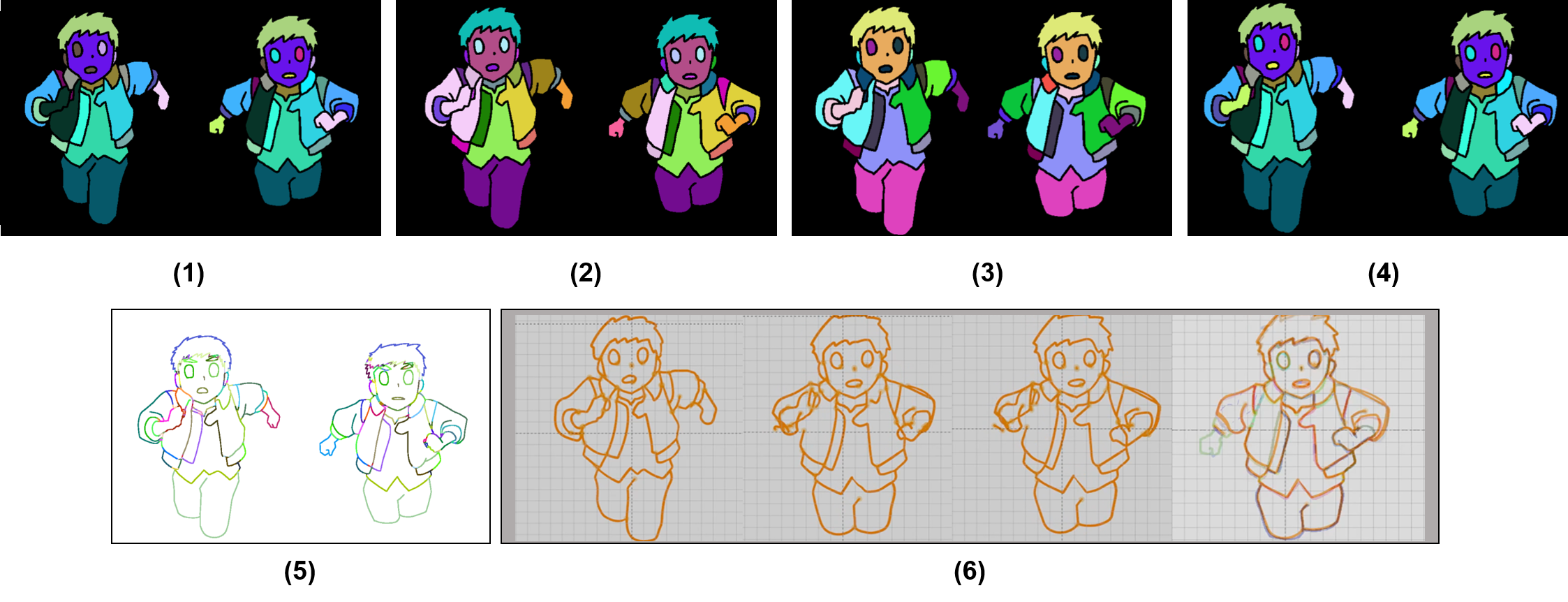}
\subcaption{$C_c$}
\label{e13}
\end{minipage}
\begin{minipage}[b]{0.52\linewidth}
\includegraphics[clip,width=90mm]{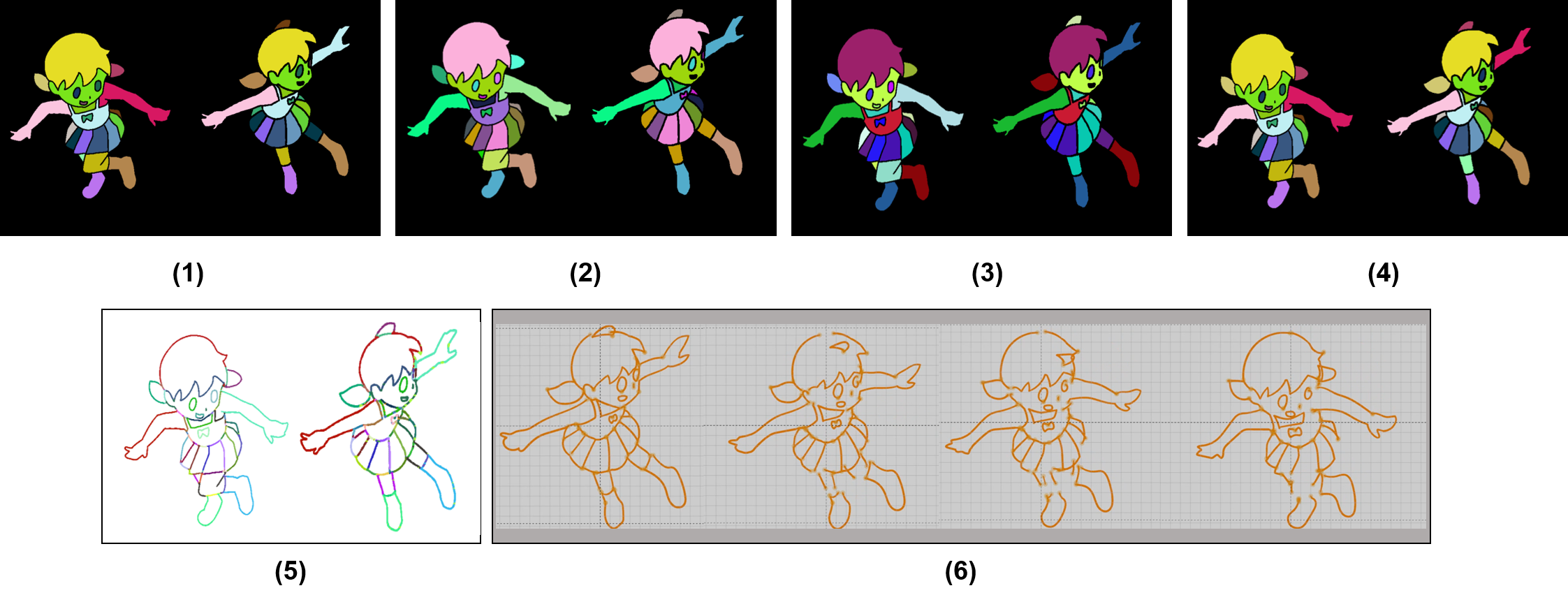}
\subcaption{$C_d$}
\label{e14}
\end{minipage}
\caption{Estimation results of closed area correspondences  using (1)~$SCD$, (2)~$SC$, and (3)~$S$ methods, as well as (4)~user corrected correspondences. Given the closed area correspondences , we estimate (5)~stroke correspondences and generate (6)~an inbetweening sequence.}
\label{e}

\end{figure*}
%+++++++++++++++++++++++++

\section{Results}

\subsection{Objective Evaluation}
Figure~\ref{e} shows the estimated results of ~$SCD$, ~$SC$, and ~$S$ methods. 
This figure also shows ~the results of user-corrected correspondences, ~the estimated stroke correspondence based on the closed area correspondences, and ~the results of automatically generated inbetweenings of $C_{a \sim d}$ sequences based on the proposed estimated stroke correspondence. 
Note that the colors of the closed areas and strokes are randomly assigned, but the pairs of closed area/strokes are given the same colors. Table~\ref{table:data_type1} and Table~\ref{table:data_type2} show the matching accuracy of closed area correspondences and the average number of user corrections. 

\begin{figure}[t]
\centerline{\includegraphics[clip,width=0.82\linewidth]{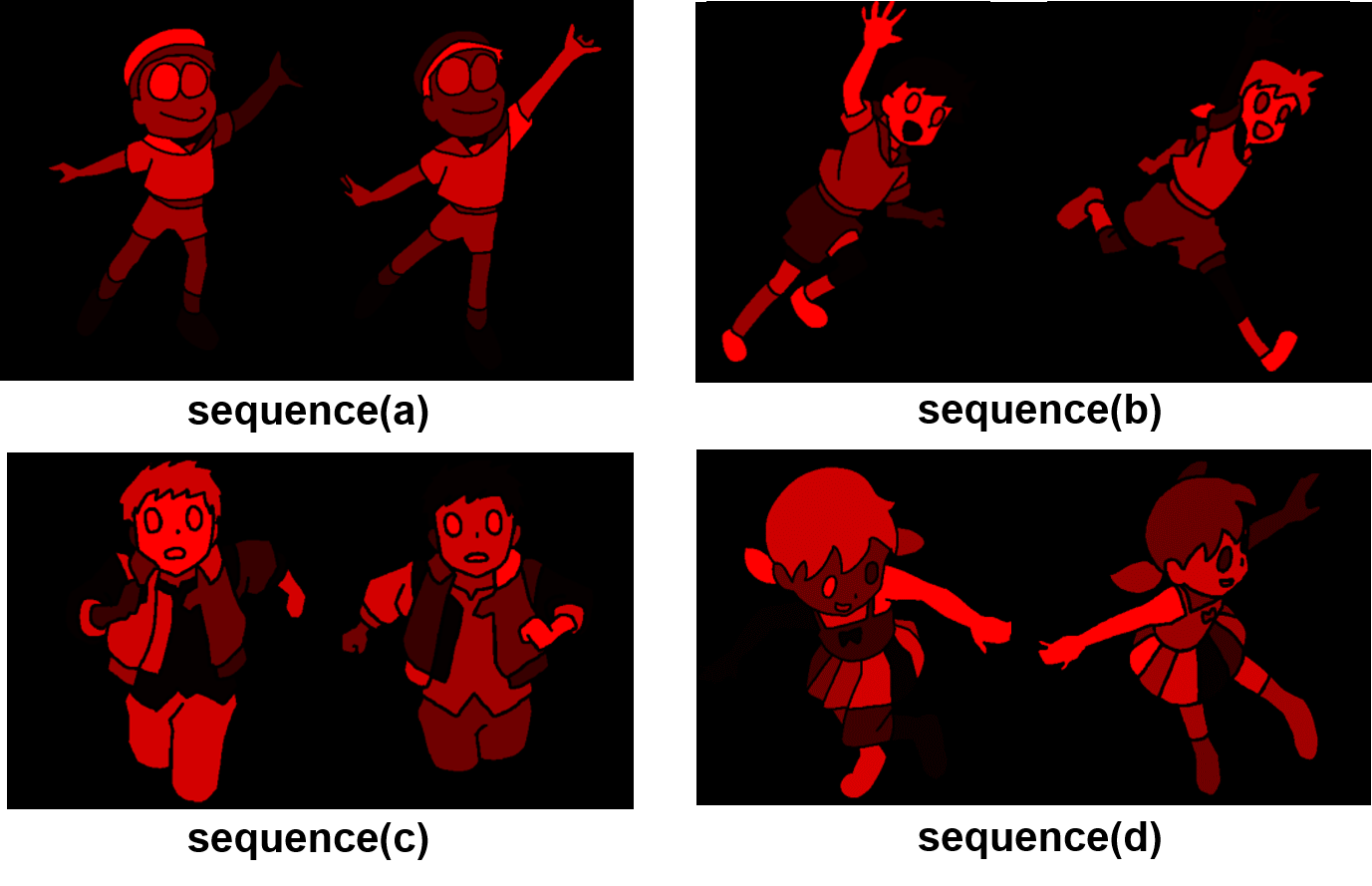}}%60mm
\caption{Depth estimation results of $C_{a \sim d}$ sequences.}
\label{depth}
\end{figure}

The shape-only approach $S$ is lower in the estimation accuracy than other methods ($SCD$  and $SC$), so the characteristics of connection relationship is needed to produce better estimations. On the other hand, from the results of $SCD$ (with the depth relationship estimation) and $SC$ (without the depth relationship estimation), we confirm that the quality of the proposed system depends on input scenes. If input keyframes have a similar depth relationship like $C_d$ (see Figure~\ref{depth}d), we can obtain plausible correspondences and reduce the number of user corrections. On the other hand, if it is difficult to determine the depth relationships (see Figure~\ref{depth}a,b,c), $SCD$ has lower accuracy than $SC$. 

\begin{table}[t]
  \caption{The percentage accuracy of closed area correspondences estimation.}
%  \caption{closed area correspondences [\%]}
   \label{table:data_type1}
    \begin{center}
     \begin{tabular}{|l|c|c|c|c|} \hline
%      Method & $C_a$ & $C_b$ & $C_c$ & $C_d$ \\ \hline \hline
    \multirow{2}{*}{} & \multicolumn{4}{c|}{Animation Cuts} \\\cline{2-5}
    & $C_a$ & $C_b$ & $C_c$ & $C_d$ \\ \hline %\hline
      $SCD$ & 86.364 &46.154&71.429&69.595\\ %\hline
      $SC$ & 86.364 &57.692&79.592&58.670\\ %\hline
      $S$ & 56.522 &57.692&59.184&41.304\\ \hline
   \end{tabular}
  \end{center}
\end{table}

%+++++++++++++++++++++++++
\begin{table}[t]
  \caption{The average number of user corrections for closed area correspondence.}
  %\caption{number of corrections [times]}
   \label{table:data_type2}
    \begin{center}
     \begin{tabular}{|l|c|c|c|c|} \hline
%     Method & $C_a$ & $C_b$ & $C_c$ & $C_d$ \\ \hline \hline
     \multirow{2}{*}{} & \multicolumn{4}{c|}{Animation Cuts} \\\cline{2-5}
     & $C_a$ & $C_b$ & $C_c$ & $C_d$ \\ \hline %\hline
      $SCD$ &3 &12 & 7 &9  \\ %\hline
      $SC$ &2 & 10 & 4 &10 \\ %\hline
      $S$ & 9& 15 & 12 &17 \\ \hline
   \end{tabular}
  \end{center}
\end{table}

%+++++++++++++++++++++++++

\begin{figure}[t]
\centerline{
\includegraphics[clip,width=0.7\linewidth]{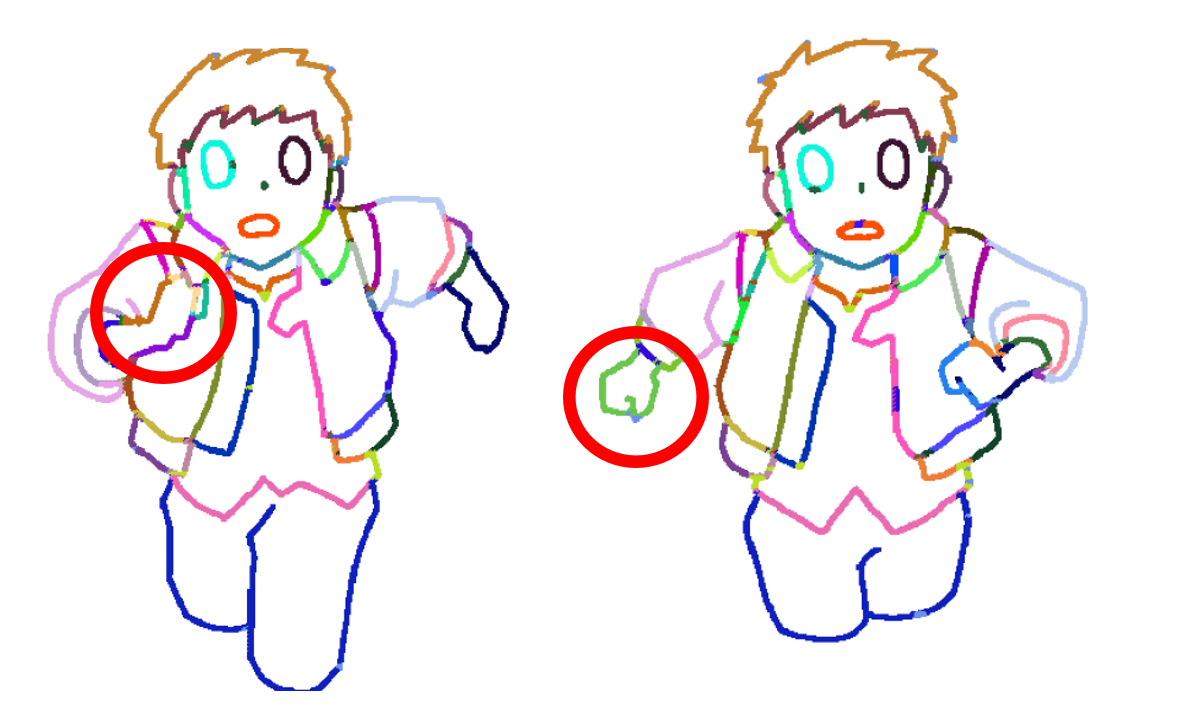}}
\caption{Failure case. Closed area and stroke correspondences between two keyframes might be changed due to self-occuluded closed areas.}
\label{missmatch}
\end{figure}

The average accuracy of stroke correspondence estimation using our methods is 59.513\% ($C_{a}$: 69.595\%, $C_{b}$: 58.670\%, $C_{c}$: 41.304\%, and $C_{d}$: 68.481\%). 
A possible reason why the estimation accuracy is insufficient ($>$ 80\%) is that the actual relationships between closed areas are changed in the input keyframes and strokes might disappear due to occlusion, so several strokes are identified as ``no correspondence,'' as shown in Figure~\ref{missmatch}.
%+++++++++++++++++++++++++

In addition, the proposed system has difficulties automatically generating plausible inbetweening frames (see Figure~\ref{e}) since the accuracy of the stroke correspondence estimation is insufficient. However, the current qualities are sufficient to automatically generate tight inbetweenings like $C_{a}$, the upper parts of $C_{b}$ \& $C_{d}$, and the lower parts of $C_{c}$.

\subsection{User Experience}
\begin{figure}[t]
\includegraphics[clip,width=\linewidth]{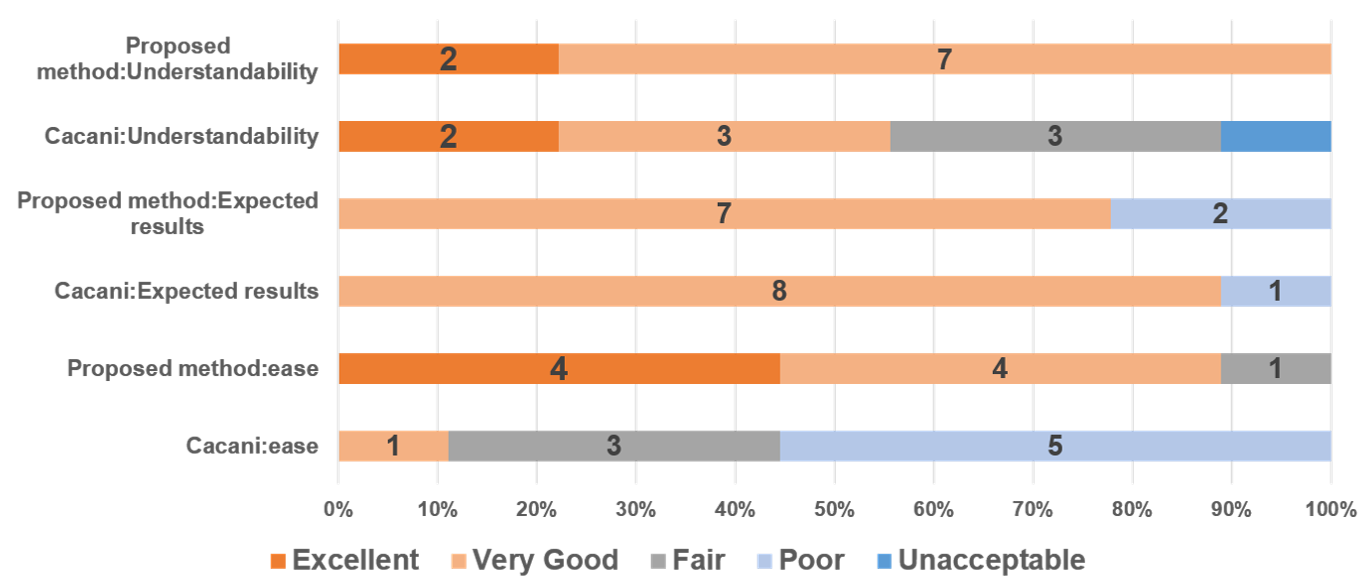}
\caption{The results of post-experiment questionnaire.}
\label{ex2}
\end{figure}

Figure~\ref{ex2} shows the post-experiment questionnaire results. We confirmed that participants who were more accustomed to draw illustrations felt more comfortable designing cartoon images with Cacani, but all participants answered that the proposed method was easier than working with Cacani. 
Regarding the ease of understanding the operation, they also answered that the proposed method was equivalent or easier to understand compared to Cacani. 
Note that the average operating time was 7 min 17 sec for Cacani and 1 min 6 sec for the proposed method. Therefore, we can conclude that the proposed method imposed fewer burdens on the participants than Cacani.

\section{Conclusion}
In this paper, we proposed a novel method to estimate stroke correspondences between two raster images with labeled closed areas. 
The proposed method allows users to manually edit the estimated results, thus improving the accuracy of the correspondences. We verified the feasibility of the proposed method using four types of sequences. In addition, the comparison study and the questionnaire suggest that the burden on the users can be reduced when using the proposed method compared to the conventional tools when determining stroke correspondence. Therefore, the proposed method of labeling closed areas is effective in estimating the stroke correspondence, especially in cases where the corresponding closed areas do not change significantly. The participants reported that estimating stroke correspondences was easier and quicker with the proposed approach.

In our prototype, we found that the accuracy drops in estimating stroke correspondence when the relationship of closed areas changes significantly between the keyframes, for example, a new closed area may be generated due to large movements of characters. If a stroke exists independently without a closed area such as wrinkles on clothes, the proposed approach will fail to calculate the stroke correspondence. In order to solve the above issues and increase estimation accuracy, we plan to split the strokes in the correct position and improve the estimation accuracy of depth relationships using deep learning based 3D human shape and pose \cite{training2019}. Note that AKAZE may be inappropriate for extracting feature points for keyframes due to less information such as color differences and shadow information compared to photographs. Therefore, the feature extraction method can be improved in future.

To improve the use of the proposed approach, we would like to develop the user interfaces for not only user correction, but also user guidance in sketch input \cite{dualface}.    
The depth estimation has low accuracy due to inaccurate extraction of the required junctions. The possible solution might be to detect closed area occluded by other closed area to improve the estimation of hidden strokes. Finally, we consider the inbetweening frames for freehand drawing of the keyframes \cite{xing15}.

\section*{Acknowledgment}
This work was supported by Grant from Tateishi Science and Technology Foundation, JSPS KAKENHI grant JP20K19845 and JP19K20316, Japan.

{\small
\bibliographystyle{IEEEtran}
\bibliography{reference}

\begin{thebibliography}{10}
\providecommand{\url}[1]{#1}
\csname url@samestyle\endcsname
\providecommand{\newblock}{\relax}
\providecommand{\bibinfo}[2]{#2}
\providecommand{\BIBentrySTDinterwordspacing}{\spaceskip=0pt\relax}
\providecommand{\BIBentryALTinterwordstretchfactor}{4}
\providecommand{\BIBentryALTinterwordspacing}{\spaceskip=\fontdimen2\font plus
\BIBentryALTinterwordstretchfactor\fontdimen3\font minus
  \fontdimen4\font\relax}
\providecommand{\BIBforeignlanguage}[2]{{%
\expandafter\ifx\csname l@#1\endcsname\relax
\typeout{** WARNING: IEEEtran.bst: No hyphenation pattern has been}%
\typeout{** loaded for the language `#1'. Using the pattern for}%
\typeout{** the default language instead.}%
\else
\language=\csname l@#1\endcsname
\fi
#2}}
\providecommand{\BIBdecl}{\relax}
\BIBdecl

\bibitem{baxter2006latent}
W.~Baxter and K.-i. Anjyo, ``Latent doodle space,'' \emph{Computer Graphics
  Forum}, vol.~25, no.~3, pp. 477--485, 2006.

\bibitem{boris2015vector}
B.~Dalstein, R.~Ronfard, and M.~van~de Panne, ``Vector graphics animation with
  time-varying topology,'' \emph{ACM Transactions on Graphics (TOG)}, vol.~34,
  no.~4, pp. 145:1--145:12, 2015.

\bibitem{di2001automatic}
F.~Di~Fiore, P.~Schaeken, K.~Elens, and F.~Van~Reeth, ``Automatic in-betweening
  in computer assisted animation by exploiting 2.5 d modelling techniques,'' in
  \emph{Proceedings Fourteenth Conference on Computer Animation (Cat. No.
  01TH8596)}.\hskip 1em plus 0.5em minus 0.4em\relax Seoul, South Korea: IEEE,
  2001, pp. 192--200.

\bibitem{fukusato2016active}
T.~Fukusato and S.~Morishima, ``Active comicing for freehand drawing
  animation,'' in \emph{Mathematical Progress in Expressive Image Synthesis
  III}.\hskip 1em plus 0.5em minus 0.4em\relax Singapore: Springer, 2016, pp.
  45--56.

\bibitem{fukusato2021view}
\BIBentryALTinterwordspacing
T.~Fukusato and A.~Maejima, ``View-dependent formulation of 2.5d cartoon
  models,'' \emph{Arxiv}, 2021. [Online]. Available:
  \url{https://arxiv.org/abs/2103.15472}
\BIBentrySTDinterwordspacing

\bibitem{fujita2017canvas}
M.~Fujita and S.~Saito, ``Hand-drawn animation with self-shaped canvas,'' in
  \emph{ACM SIGGRAPH 2017 Posters}.\hskip 1em plus 0.5em minus 0.4em\relax New
  York, NY, USA: ACM, 2017, pp. 5:1--5:2.

\bibitem{Furusawa2014QR}
C.~Furusawa, T.~Fukusato, N.~Okada, T.~Hirai, and S.~Morishima, ``Quasi 3d
  rotation for hand-drawn characters,'' in \emph{ACM SIGGRAPH 2014
  Posters}.\hskip 1em plus 0.5em minus 0.4em\relax New York, NY, USA: ACM,
  2014, pp. 12:1--12:1.

\bibitem{Marek2018}
M.~Dvoro\v{z}n\'{a}k, W.~Li, V.~G. Kim, and D.~S\'{y}kora, ``Toonsynth:
  Example-based synthesis of hand-colored cartoon animations,'' \emph{ACM
  Transactions on Graphics (TOG)}, vol.~37, no.~4, pp. 167:1--167:11, 2018.

\bibitem{sederberg1992physically}
T.~W. Sederberg and E.~Greenwood, ``A physically based approach to 2--d shape
  blending,'' \emph{ACM SIGGRAPH Computer Graphics}, vol.~26, no.~2, pp.
  25--34, 1992.

\bibitem{sederberg19932}
T.~W. Sederberg, P.~Gao, G.~Wang, and H.~Mu, ``2-d shape blending: an intrinsic
  solution to the vertex path problem,'' in \emph{Proceedings of the 20th
  Annual Conference on Computer Graphics and Interactive Techniques},
  vol.~93.\hskip 1em plus 0.5em minus 0.4em\relax New York, NY, USA: ACM, 1993,
  pp. 15--18.

\bibitem{yang2017context}
W.~Yang, ``Context-aware computer aided inbetweening,'' \emph{IEEE Transactions
  on Visualization and Computer Graphics (TVCG)}, vol.~24, no.~2, pp.
  1049--1062, 2017.

\bibitem{cacani2021}
{CACANi Private Ltd}, ``Cacani: 2d animation \& inbetween software,''
  \url{https://cacani.sg/}, 2021.

\bibitem{burtnyk1975computer}
N.~Burtnyk and M.~Wein, ``Computer animation of free form images,'' in
  \emph{Proceedings of the 2nd Annual Conference on Computer Graphics and
  Interactive Techniques}.\hskip 1em plus 0.5em minus 0.4em\relax New York, NY,
  USA: ACM, 1975, pp. 78--80.

\bibitem{reeves1981inbetweening}
W.~T. Reeves, ``Inbetweening for computer animation utilizing moving point
  constraints,'' in \emph{Proceedings of the 8th Annual Conference on Computer
  Graphics and Interactive Techniques}, vol.~15, no.~3.\hskip 1em plus 0.5em
  minus 0.4em\relax New York, NY, USA: ACM, 1981, pp. 263--269.

\bibitem{Juan2006reusing}
C.~N. de~Juan and B.~Bodenheimer, ``Re-using traditional animation: Methods for
  semi-automatic segmentation and inbetweening,'' in \emph{Proceedings of the
  2006 ACM SIGGRAPH/Eurographics Symposium on Computer Animation (SCA
  '06)}.\hskip 1em plus 0.5em minus 0.4em\relax Goslar, DEU: Eurographics
  Association, 2006, p. 223–232.

\bibitem{baxter2009compatible}
W.~Baxter, P.~Barla, and K.-i. Anjyo, ``Compatible embedding for 2d shape
  animation,'' \emph{IEEE Transactions on Visualization and Computer Graphics
  (TVCG)}, vol.~15, no.~5, pp. 867--879, 2009.

\bibitem{fujita2019}
M.~Fujita and S.~Saito, ``Matching of interkeyframe strokes considering
  pairwise-constraints using self-shaped canvas,'' in \emph{Proceedings of the
  81-th National Convention of IPSJ}, vol. 2019, no.~1, 2019, pp. 165--166.

\bibitem{white2010betweenit}
B.~Whited, G.~Noris, M.~Simmons, R.~W. Sumner, M.~Gross, and J.~Rossignac,
  ``Betweenit: An interactive tool for tight inbetweening,'' \emph{Computer
  Graphics Forum}, vol.~29, no.~2, pp. 605--614, 2010.

\bibitem{yang2018ftp}
W.~Yang, H.-S. Seah, Q.~Chen, H.-Z. Liew, and D.~S{\`y}kora, ``Ftp-sc: Fuzzy
  topology preserving stroke correspondence,'' \emph{Computer Graphics Forum},
  vol.~37, no.~8, pp. 125--135, 2018.

\bibitem{yamaguchi2015region}
S.~Yamaguchi, T.~Kato, T.~Fukusato, C.~Furusawa, and S.~Morishima,
  ``Region-based painting style transfer,'' in \emph{SIGGRAPH Asia 2015
  Technical Briefs}.\hskip 1em plus 0.5em minus 0.4em\relax New York, NY, USA:
  ACM, 2015, pp. 8:1--8:4.

\bibitem{sato2014reference}
K.~Sato, Y.~Matsui, T.~Yamasaki, and K.~Aizawa, ``Reference-based manga
  colorization by graph correspondence using quadratic programming,'' in
  \emph{SIGGRAPH Asia 2014 Technical Briefs}.\hskip 1em plus 0.5em minus
  0.4em\relax New York, NY, USA: ACM, 2014, pp. 15:1--15:4.

\bibitem{maejima2019graph}
A.~Maejima, H.~Kubo, T.~Funatomi, T.~Yotsukura, S.~Nakamura, and Y.~Mukaigawa,
  ``Graph matching based anime colorization with multiple references,'' in
  \emph{ACM SIGGRAPH 2019 Posters}.\hskip 1em plus 0.5em minus 0.4em\relax New
  York, NY, USA: ACM, 2019, pp. 13:1--13:2.

\bibitem{akaze}
P.~Alcantarilla, J.~Nuevo, and A.~Bartoli, ``Fast explicit diffusion for
  accelerated features in nonlinear scale spaces,'' in \emph{Proceedings of the
  British Machine Vision Conference}.\hskip 1em plus 0.5em minus 0.4em\relax
  BMVA Press, 2013, pp. 13:1--13:11.

\bibitem{training2019}
H.~{Xie}, A.~{Watatani}, and K.~{Miyata}, ``Visual feedback for core training
  with 3d human shape and pose,'' in \emph{2019 Nicograph International
  (NicoInt)}, 2019, pp. 49--56.

\bibitem{dualface}
\BIBentryALTinterwordspacing
Z.~Huang, Y.~Peng, T.~Hibino, C.~Zhao, H.~Xie, T.~Fukusato, and K.~Miyata,
  ``dualface: Two-stage drawing guidance for freehand portrait sketching,''
  \emph{CoRR}, vol. abs/2104.12297, 2021. [Online]. Available:
  \url{https://arxiv.org/abs/2104.12297}
\BIBentrySTDinterwordspacing

\bibitem{xing15}
\BIBentryALTinterwordspacing
J.~Xing, L.-Y. Wei, T.~Shiratori, and K.~Yatani, ``Autocomplete hand-drawn
  animations,'' \emph{ACM Trans. Graph.}, vol.~34, no.~6, Oct. 2015. [Online].
  Available: \url{https://doi.org/10.1145/2816795.2818079}
\BIBentrySTDinterwordspacing

\end{thebibliography}
}

\end{document}